\documentclass[12pt]{article}
\usepackage[english]{babel}
\usepackage{xcolor}

\usepackage{scicite}
\usepackage{times}

\usepackage{graphicx}

\newenvironment{sciabstract}{%
\begin{quote} \bf}
{\end{quote}}

\title{High-rate Generation and State Tomography of Non-Gaussian Quantum States for Ultra-fast Clock Frequency Quantum Processors}

\author
{Akito Kawasaki,$^{1,\dagger}$ Ryuhoh Ide,$^{1}$ Hector Brunel,$^{1,2}$ Takumi Suzuki,$^{1}$\\
Rajveer Nehra,$^{1,3,4}$ Katsuki Nakashima,$^{1}$ Takahiro Kashiwazaki,$^{5}$ Asuka Inoue,$^{5}$\\
Takeshi Umeki,$^{5}$ Fumihiro China,$^{6}$ Masahiro Yabuno,$^{6}$ Shigehito Miki,$^{6}$\\
Hirotaka Terai,$^{6}$ Taichi Yamashima,$^{1}$ Atsushi Sakaguchi,$^{7}$ Kan Takase,$^{1,7}$\\
Mamoru Endo,$^{1,7}$ Warit Asavanant,$^{1,7,\ast}$ Akira Furusawa$^{1,7,\ast\ast}$\\
\\
\normalsize{$^{1}$Department of Applied Physics, School of Engineering, The University of Tokyo,}\\
\normalsize{7-3-1 Hongo, Bunkyo-ku, Tokyo 113-8656, Japan}\\
\normalsize{$^{2}$Department of Physics, Ecole Normale Sup\'{e}rieure,}\\
\normalsize{24, rue Lhomond, Paris 75005, France}\\
\normalsize{$^{3}$Department of Electrical and Computer Engineering, University of Massachusetts Amherst,}\\
\normalsize{Amherst, Massachusetts 01003, USA}\\
\normalsize{$^{4}$Department of Physics, University of Massachusetts Amherst,}\\
\normalsize{Amherst, Massachusetts 01003, USA}\\
\normalsize{$^{5}$NTT Device Technology Labs, NTT Corporation,}\\
\normalsize{3-1 Morinosato Wakamiya, Atsugi, Kanagawa 243-0198, Japan}\\
\normalsize{$^{6}$Advanced ICR Research Institute, National Institute of Information and}\\
\normalsize{Communications Technology, 588-2 Iwaoka, Nishi, Kobe 651-2492, Japan}\\
\normalsize{$^{7}$Optical Quantum Computing Research Team, RIKEN Center for Quantum Computing,}\\
\normalsize{ 2-1 Hirosawa, Wako, Saitama 351-0198, Japan}\\
\normalsize{To whom correspondence should be addressed; $^{\dagger}$E-mail: kawasaki@alice.t.u-tokyo.ac.jp}\\
\normalsize{$^{\ast}$E-mail: warit@alice.t.u-tokyo.ac.jp $^{\ast\ast}$E-mail: akiraf@ap.t.u-tokyo.ac.jp}
}

\date{}

\begin{document} 


\baselineskip24pt


\maketitle


\begin{sciabstract}
Quantum information processors greatly benefit from  high clock frequency to fully harnessing the quantum advantages before they get washed out by the decoherence.  In this pursuit, all-optical systems offer unique advantages due to their inherent 100 THz carrier frequency, permitting one to develop THz clock frequency processors. In practice, the bandwidth of the quantum light sources and the measurement devices has been limited to the MHz range and the generation rate of nonclassical states to kHz order---a tiny fraction of what can be achieved. In this work, we go beyond this limitation by utilizing optical parametric amplifier (OPA) as a squeezed-light source and optical phase-sensitive amplifiers (PSA) to realize high-rate generation of broadband non-Gaussian states and their quantum tomography. Our state generation and measurement system consists of a 6-THz squeezed-light source, a 6-THz PSA, and a 66-GHz homodyne detector. With this system, we have successfully demonstrated non-Gaussian state generation at a 0.9 MHz rate---almost three orders of magnitude higher than the current state-of-the-art experiments---with a sub-nanosecond wave packet using continuous-wave laser. The performance is constrained only by the superconducting detector's jitter which currently limits the usable bandwidth of the squeezed light to 1 GHz, rather than the optical and electronic systems. Therefore, if we can overcome the limitation of the timing jitter of superconducting detector, non-Gaussian state generation and detection at GHz rate, or even THz rate, for optical quantum processors might be possible with OPAs.

\end{sciabstract}

% In setting up this template for *Science* papers, we've used both
% the \section* command and the \paragraph* command for topical
% divisions.  Which you use will of course depend on the type of paper
% you're writing.  Review Articles tend to have displayed headings, for
% which \section* is more appropriate; Research Articles, when they have
% formal topical divisions at all, tend to signal them with bold text
% that runs into the paragraph, for which \paragraph* is the right
% choice.  Either way, use the asterisk (*) modifier, as shown, to

\section*{Introduction}
\begin{figure*}[htb]
 \includegraphics[width=\textwidth]{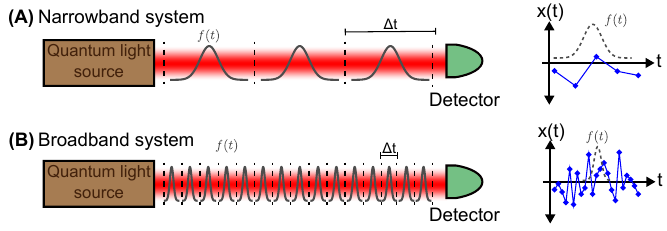}%
 \caption{Bandwidth and clock frequency. (A) Narrowband system. (B) Broadband system. $f(t)$, the shape of the wave packet; $x(t)$ temporal signal from the detector. The frequency bandwidth of the quantum light source limits the size $\Delta t$ of the wave packet we can select. The bandwidth of the detector limits the sampling rate of $x(t)$. This sampling rate has to be sufficiently high compare to the time scale of $f(t)$.\label{fig:schematic}}
 \end{figure*}
With the recent advances in quantum technologies, various applications such as quantum computation and quantum communication \cite{nielsen00} are expected to be the next technological leap. Clock frequency or the speed of the quantum information processors is one of the most important factor determining the scalability and applications of the quantum technologies; if the generation or manipulation of quantum states is slow, the nonclassicality may dissipate due to the decoherence before any practical applications can be realised. The ultimate physical limit of the clock frequency is set by the carrier frequency of the physical system. Note that this carrier frequency does not directly translate to the clock frequency as the quantum information is usually encoded in frequency around the carrier frequency, meaning that the clock frequency is determined by the bandwidth. Figure \ref{fig:schematic} visualizes this relationship. This is similar to how optical telecommunication or radio encode classical information in the amplitude or frequency modulation, not at the carrier frequency, and the bandwidth of the modulation determines how fast information can be sent. In this regard, Optical systems are one of the most promising platform for building the utility-scale quantum processors. The carrier frequency of optical system is a few hundred THz meaning that, in principle, high-speed state generation and measurements can be achieved. Moreover, by utilizing the rich degree of freedom, various quantum resourcess multiplexing techniques have been achieved \cite{10.1063/9780735424074,Asavanant373,Larsen369,PhysRevLett.112.120505}, demonstrating potential for large-scale quantum applications with minimal physical resources.

Despite the capability, the current clock frequency of the optical quantum processors is far from THz. As almost every quantum applications can be broken down to state preparation, manipulation, and detection, the bandwidth of these operations determines the clock frequency. The bandwidth of the nonclassical light source called squeezed light, a ubiquitous light source in state generation, and the bandwidth of the homodyne detectors is one of the main limiting factors. In particular, despite demonstrations of various state generations and applications \cite{10.1063/9780735424074,doi:10.1063/1.5100160}, the generation rate of the non-Gaussian state remains on the kHz order and the bandwidth of the homodyne detector is limited to around hundred MHz \cite{Asavanant:17,Takase:22}. Note that although the bandwidth of the squeezed light typically generated with optical parametric oscillators (OPOs) can be MHz to GHz, the optical state generation uses heralding process \cite{10.1063/9780735424074} which reduces the generation rate. To overcome this, we have seen the usages of optical parametric amplifier (OPA) as a squeezed light source and a phase sensitive amplifier (PSA) for broadband measurements \cite{Shaked2018,doi:10.1063/1.5142437,kashiwazaki,10.1063/5.0137641,doi:10.1126/science.abo6213}. OPA has been shown to have the bandwidth on the THz order \cite{doi:10.1063/1.5142437,doi:10.1126/science.abo6213,OptLett.47.1506}. Among various demonstrations, the measurement of the 43-GHz single-mode squeezed light with OPA is the current most advanced result \cite{10.1063/5.0137641}. For applications such as quantum computation, however, we need to be able to achieve high-rate generation and broadband measurement of non-Gaussian states or multimode entangled states as the single-mode squeezed states do not have non-Gaussianity or entanglement required in achieving quantum advantage \cite{PhysRevLett.88.097904,PhysRevLett.82.1784}. An example of a task that is heavily affected by the low generation rate is the physical realization of the logical qubit state such as Gottesman-Kitaev-Preskill (GKP) qubit \cite{PhysRevA.64.012310,PhysRevA.101.032315} for fault-tolerant quantum computation. It requires interference of multiple non-Gaussian states \cite{Vasconcelos:10,PhysRevA.97.022341,2024arXiv240107287T} and although it has been recently demonstrated for a single-step interference \cite{doi:10.1126/science.adk7560}, multiple-step interference is essential in achieving GKP with required quality for fault-tolerant quantum computation. Because the overall generation rate reduces exponentially with the interference steps, generation of optical GKP state is experimentally challenging with current kHz generation rate.

In this paper, we realize the high-rate generation and homodyne measurement of non-Gaussian states. To demonstrate the versatility of our system, we perform full quantum-state tomography of the generated state and verify its quantum non-Gaussianity through the Wigner negativity. The tomography results show strong nonclassicality of the generated states. The full bandwidth of the state generation and measurement is 66 GHz, limited by electronics of our detection system. Due to the jitter of the photon detector, we limit the bandwidth down to 1 GHz, resulting in the generation rate of about 0.9 MHz. Even then, both the bandwidth and the generation rate are two orders of magnitude larger than the current state-of-the-art experiments \cite{Asavanant:17,Takase:22,Melalkia:22}. There are no other limiting factor besides the jitter of the photon detector \cite{supp}, meaning that in principle, GHz generation rate is experimentally feasible, leading to enhanced clock rates of optical processors. Recent theoretical work has shown that it is possible to use OPA at room temperature as a photon detector \cite{PRXQuantum.4.010333}, OPA might also hold a key to removing limitation imposed by the timing jitter of superconducting detector. As Gaussian states generation and measurement allows only classically simulable quantum computation \cite{PhysRevLett.88.097904}, this work opens a new path to genuine high-speed optical quantum applications. We believe that state preparation and measurement using PSA will become the next generation optical quantum technology, replacing the typical squeezed light sources and homodyne detectors.

\section*{Bandwidth and quantum state}
Optical quantum state is described by the annihilation operator $\hat{a}$ whose classical analogue is the complex electric field. If we have a quantum state with the temporal mode $f(t)$, the annihilation operator of this mode is $\hat{a}_f=\int\textrm{d}t\,f(t)\hat{a}(t)$, where $\hat{a}(t)$ is the time-dependent annihilation operator. We can also use Fourier transform to rewrite this in frequency domain as $\hat{a}_f=\int\textrm{d}\omega\tilde{f}(\omega)\hat{\tilde{a}}(\omega)$, where $\tilde{f}(\omega)$ is the Fourier transform of $f(t)$ and $\hat{\tilde{a}}(\omega)$ is the annihilation operator of single frequency mode $\omega$ ($\omega=0$ corresponds to carrier frequency). Thus, if we want $f(t)$ to be narrow so that the clock frequency is high, our light source and the detector have to be sufficiently broad to correctly generate and measured all frequency component of $\tilde{f}(\omega)$. The requirement of the broadband light source and the homodyne measurement for quantum state in temporal wave packet is visualized in Fig.\ \ref{fig:schematic}.

Although $\hat{a}$ can be used to describe the state, the actual observable that is measured is the quadrature operator $\hat{x}$ and $\hat{p}$ satisfying $[\hat{x},\hat{p}]=i$ ($\hbar=1$). We can write the quadrature of the mode $f(t)$ as $\hat{x}_f=\int\textrm{d}t\,f(t)\hat{x}(t)$, where $\hat{x}(t)$ is the quadrature values at time $t$. Thus, for the homodyne detector to be able to measure correct $\hat{x}_f$, it must have sufficient bandwidth to extract the value of the varying $\hat{x}(t)$ on at least the time scale faster than the width of $f(t)$.

\section*{Experimental setup}
\begin{figure*}[htbp]
 \includegraphics[width=\textwidth]{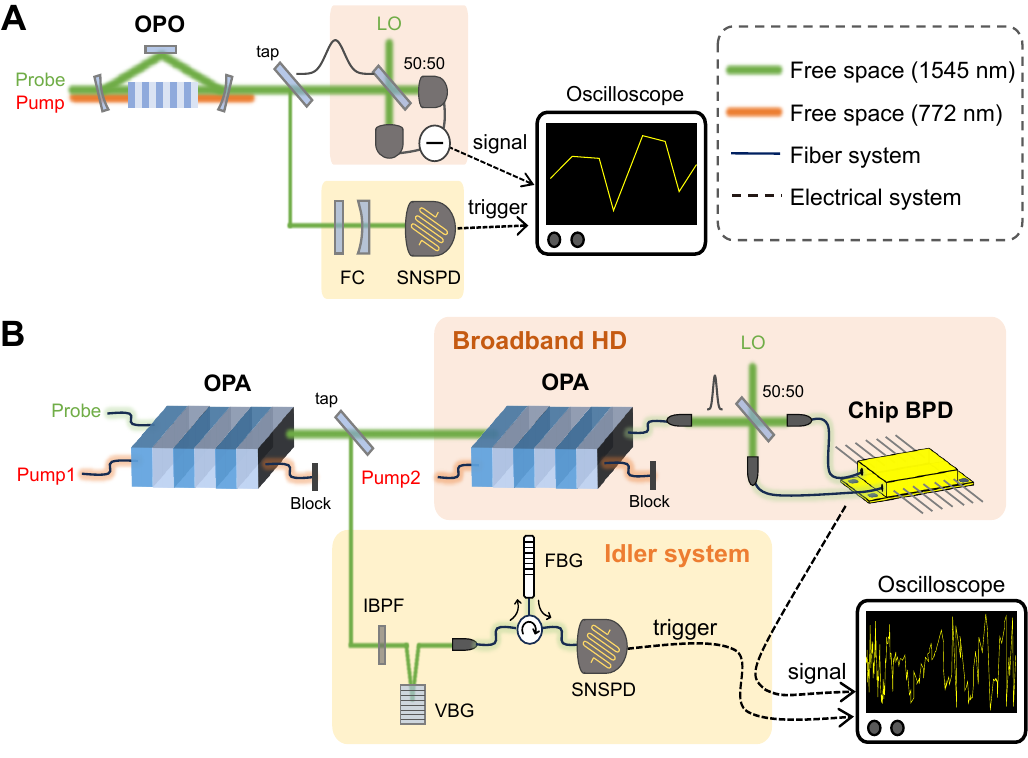}%
 \caption{\textbf{Comparison between typical narrowband state generation system and broadband state generation system in this work} (A) Narrowband system based on optical parametric oscillator. (B) Broadband state generation and measurement based on optical parametric amplifier. OPO, Optical parametric oscillator; LO, local oscillator; FC, Filter cavity; SNSPD, Superconducting nanostrip photon detector; OPA, Optical parametric amplifier; HD, Homodyne detector; BPD, Balanced photodetector; IBPF, Interference bandpass filter; VBG, Volume Bragg grating; FBG, Fiber Bragg grating. See supplementary for detailed parameters.\label{fig:exp}}
 \end{figure*}
The non-Gaussian state is generated by tapping a portion of light from the squeezed light source and send it to the superconducting nanostrip photon detector (SNSPD). This method is called photon subtraction \cite{PhysRevA.55.3184} and it is basic method used to generate approximation of Schr{\"o}dinger cat states \cite{Ourjoumtsev83,PhysRevLett.97.083604,Wakui:07,Asavanant:17}.
Figure \ref{fig:exp} shows the experimental setup. As a reference, we also show the setup of typical narrowband continuous-wave experimental setup for comparison in Fig.\ \ref{fig:exp}(a). In narrowband experiment, the squeezed light is generated by the OPO and the bandwidth is limited by the linewidth of the cavity to be below GHz order. As the time scale in the narrowband experiment is much larger than the timing jitter of SNSPD, the effect of the timing jitter is negligible. The filtering cavity is used for selecting the resonance peak of the OPO to perform the photon subtraction as narrowband homodyne detector cannot detect the signals in all cavity modes. Figure \ref{fig:exp}(b) shows the broadband setup used in this experiment. The state is generated and evaluated at the telecommunication wavelength at 1545 nm using continuous-wave laser. Instead of OPO, we are using OPA here which has the frequency bandwidth of 6 THz \cite{kashiwazaki}, much broader than the narrowband experiment. There are two OPAs used in this experiments. The first OPA is used as a squeezed light source and the second OPA is used as a PSA. Although the bandwidth of the initial squeezed light is of THz level, because the timing jitter of the SNSPD used in this experiment \cite{supp} is about 130 ps, its effects in non-negligible and we add an optical filter to limit the detection bandwidth to about 1 GHz. The frequency bandwidth of idler determines both the size of the wave packet and the generation rate. The frequency bandwidth on the idler path determine the time correlation of photons in the idler path. When the photon is detected, the next photon detection is unlikely to occur in that vicinity; we have to wait roughly the inverse of the frequency bandwidth to have a chance of detecting a photon again. The generated non-Gaussian state is verified by commercially-available 70-GHz homodyne detector after the amplification using the PSA. The amplification via PSA allows us to use lossy broadband homodyne detector to measure the quadrature of the non-Gaussian state \cite{Shaked2018,10.1063/5.0137641}. This is in contrast to homodyne detection in the typical setup which has bandwidth below hundreds of MHz due to the requirement of the quantum efficiency. See Ref.\ \cite{supp} for the details of the experimental setup.

\section*{Experimental results}
\begin{figure*}[htb]
\centering
 \includegraphics[width=0.8\textwidth]{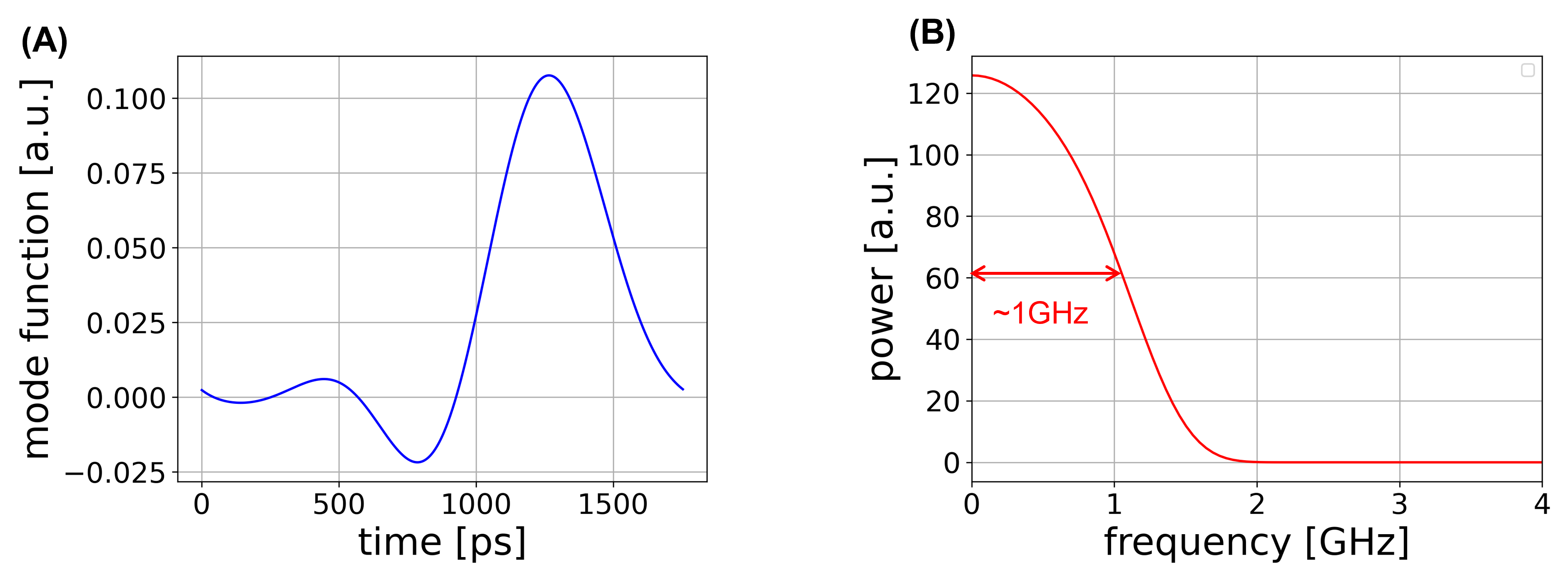}%
 \caption{Wave packet of the experimentally generated state. (A) Temporal shape obtained from principle component analysis of the measured quadrature data. (B) Amplitude square of the Fourier transform of the temporal wave packet.\label{fig:wave}}
 \end{figure*}
 \begin{figure*}[htb]
 \centering
 \includegraphics[width=0.8\textwidth]{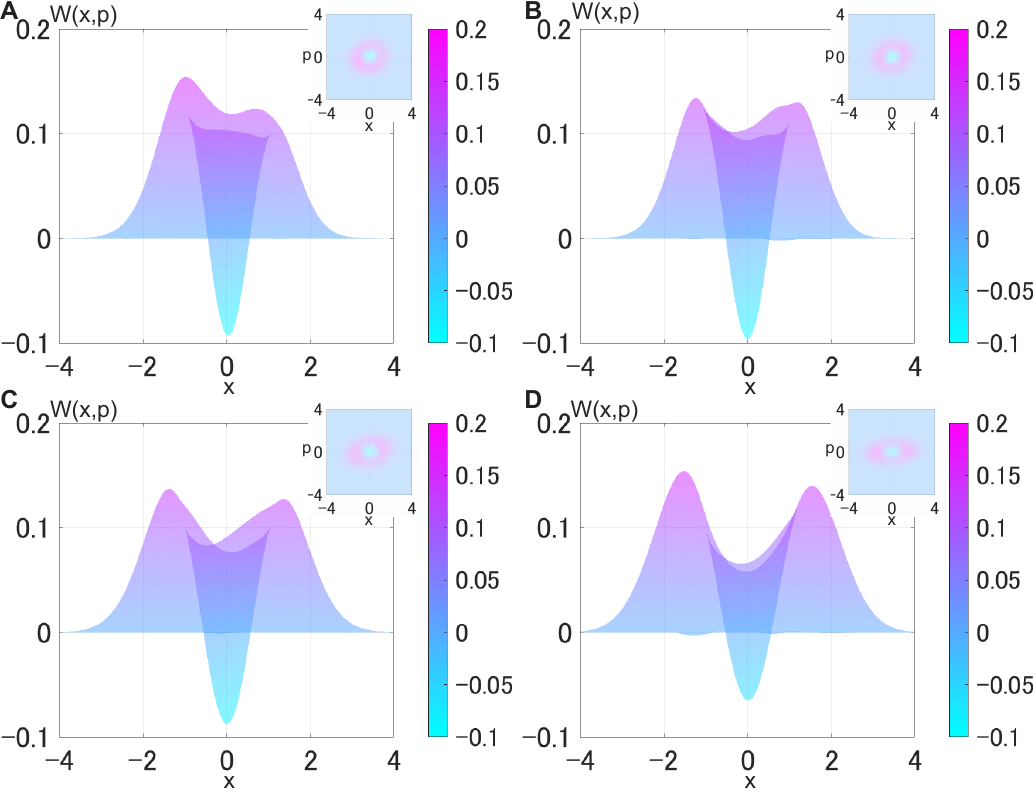}%
 \caption{\textbf{Wigner function of the reconstructed states for various pump power.} (A) 1 mW. (B) 3 mW. (C) 10 mW. (D) 25 mW. The insets are contour of the Wigner functions.\label{fig:Wigner}}
\end{figure*}

Figure \ref{fig:wave} shows the shape of the temporal wave packet of the generated state and its Fourier transform. This wave packet is determined by the principle component analysis of the collected quadrature values \cite{PhysRevLett.111.213602}. The shape of the wave packet is the expected shape and the bandwidth is 1 GHz as predicted by the spectral property of the optical filter \cite{supp}. This wave packet is, to our knowledge, the most broadband among the non-Gaussian state generated and tomographed in CW regime. We vary the pump power for the squeezer OPA as 1 mW, 3 mW, 10 mW, and 25 mW. The generation rate for each pump power is 28 kHz, 88 kHz, 310 kHz, and 0.9 MHz, respectively. Compare to the typical cat state generation with similar parameters \cite{Asavanant:17}, this is about two-order of magnitude higher. More detailed discussions is given in the supplementary \cite{supp}.

Figure \ref{fig:Wigner} shows the tomography results of the state generated state for various squeezing levels. The quantum tomography is implemented using maximum likelihood method \cite{Lvovsky_2004}. We observe negative region of Wigner functions for all the pump power used here. The negativity values of the origin (Wigner negativity) of each state at each pump power is $-0.090\pm0.004$, $-0.097\pm0.005$, $-0.088\pm0.005$, and $-0.066\pm0.006$, respectively.

\begin{figure*}[htb]
 \includegraphics[width=\textwidth]{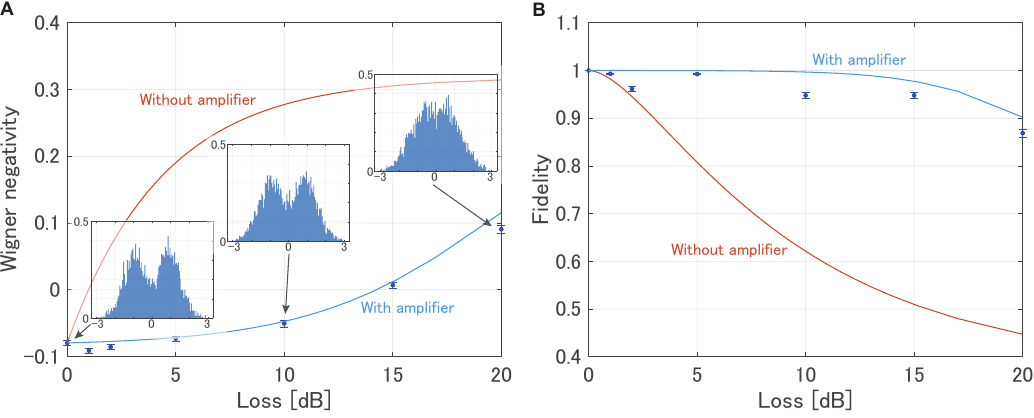}%
 \caption{\textbf{Effect of additional loss after PSA on state tomography.} (A) Dependence of Wigner negativities on additional loss. (B) Dependence between the additional loss and the fidelity to state without additional loss. We also plot the theoretical prediction for the case where same additional loss is added, but we do not use the second OPA as an amplifier. The theoretical line for the both with and without PSA cases are plotted using the experimental parameters of the setup. The insets in (A) show the quadrature distributions at antisqueezing phase for each loss. Although the dip near 0 becomes less resolvable with more loss, the non-Gaussian feature remains even with additional loss of 20 dB.\label{fig:PSA}}
\end{figure*}

To verify that the PSA allows us to correctly measure the non-Gaussian state, we also test the loss tolerance of the amplified signal. To do this, we add additional loss after the second OPA. Figure \ref{fig:PSA} shows the experimental results. First, we observe in Fig.\ \ref{fig:PSA}(A) that even with additional loss, the negativity of the reconstructed states remained even with almost 15 dB loss. Even at 20 dB loss, although the negativity is no longer present, the non-Gaussian characteristic of the quadrature distribution is still apparent. Moreover, the value of the negativities are much lower than the case without the PSA. This suggested that after the PSA, the quadrature values can be considered as classical values. We also compare the fidelity of the reconstructed state when subjected to different loss after PSA. This is shown in Fig.\ \ref{fig:PSA}(B). We observe that the fidelity agrees well with theoretical prediction \cite{supp} and remain close to 1 for the loss up to 10 dB. Comparing to the case without PSA, this result highlights the fact that optical preamplification using PSA allows us to correctly retrieve highly nonclassical quantum information.

\section*{Discussions and conclusions}
State preparation, manipulation, and measurement are key components of quantum technologies. The bandwidth of these components limits the clock frequency of the system. In this work, we have utilized the broadband OPA to overcome the current technological limitation in both generation rate and measurement. The key point here is that OPA can be used as both squeezed-light source and PSA. Moreover, it is theoretically shown that as a PSA, OPA can also be used in the classical channel of the feedforward operation which will enable all-optical quantum teleportation \cite{Ralph:99}. Thus, the remaining factor that is limiting the clock frequency of the optical quantum processors is the timing jitter of the superconducting detector. One method to remove this is to use the timing jitter remover \cite{PhysRevA.105.043714}. There is a proposal that theoretically shown that OPA can be used to implement room-temperature high-speed photon number resolving detector (PNRD) \cite{PRXQuantum.4.010333}. As the timing jitter in this approach will be negligible, replacing superconducting detector with OPA could be a possible way to remove the remaining technological limitation of the clock frequency. As the frequency bandwidth of the OPA is THz, if we can do PNRD using OPA, GHz (or even THz) generation rate of non-Gaussian state might be possible. Hence, our experiment is the first is the first step toward the possibility of genuinely room temperature optical quantum computation using only OPA and telecommunication optical devices. OPA is the next generation device that will supersede various optical quantum technology and for high clock frequency quantum processors, OPA is all you need.

\section*{Acknowledgments}
\textbf{Funding:}This work was partly supported by Japan Science and Technology (JST) Agency (Moonshot R \& D) Grant No. JPMJMS2064 and JPMJMS2066, UTokyo Foundation, and donations from Nichia Corporation of Japan. W.A. acknowledges the funding from Japan Society for the Promotion of Science (JSPS) KAKENHI (No. 23K13040). K.T. acknowledges the funding from JSPS KAKENHI (No. 23K13038, 22K20351).  M. E. acknowledges the funding from JST (JPMJPR2254). W.A., K.T, and M.E. acknowledge supports from Research Foundation for OptoScience and Technology. A.K., R.I, and T.S. acknowledge financial support from The Forefront Physics and Mathematics Program to Drive Transformation (FoPM). A.K. acknowledges financial support from Leadership Development Program for Ph.D. (LDPP), the University of Tokyo.
\textbf{Author contributions:} A.K. and W.A. conceived the experiment. A.K. and R.I. lead the experiment with supervision from W.A., K.T., A.S., M.E., and A.F.. A.K., R.I., H.B., T.S., and W.A. build the experimental setup. A.K., R.I., and K.N. collect the experimental data. A.K. and W.A. analyze the data. W.A. and A.K. visualize the data. T.K., A.I., and T.U. provide the OPA used in the experiment. F.C., M.Y., S.M., and H.T. provide the SNSPD used in this experiment. R.N., T.Y. and A.S. provide valuable advises and discussions regarding OPA and experimental system. W.A. writes the manuscript and A.K. writes the supplementary material with assistance of all the other authors.\textbf{Competing interest:} The authors declare no competing interests. \textbf{Data and materials availability:} All data used in the experiment will be available online \cite{data_online}.

\section*{Supplementary materials}
Materials and Methods\\
Figure S1\\
Reference (35)-(36)

\end{document}